\documentclass[10pt,notitlepage,a4paper,aps,prd,onecolumn,superscriptaddress,nofootinbib,groupedaddress]{revtex4-1}
\usepackage{array}

\usepackage[dvipsnames]{xcolor}

\usepackage[pdfencoding=auto, psdextra]{hyperref}
\usepackage[T1]{fontenc}
\usepackage[utf8]{inputenc}
\usepackage[english]{babel}

\usepackage[mathscr]{euscript}
\usepackage{caption}
\usepackage{subcaption}
\hypersetup{
colorlinks = true,
linkcolor = blue,
filecolor = magenta,
urlcolor = blue,
pdftitle = {TLSs interaction},
pdfpagemode = FullScreen,
citecolor = blue}
\usepackage{physics}

\usepackage{caption}

\usepackage{indentfirst}

\usepackage[pdftex]{graphicx}

\usepackage{amsmath}
\allowdisplaybreaks  % para que pueda partir fórmulas que ocupan más de una línea, necesita el paquete anterior
\usepackage{amssymb} % para cargar algunos símbolos como \blacksquare y \square
\usepackage{amsfonts} % para cargar algunas fuentes en estilo matemático
\usepackage{enumerate}

\usepackage{fancyhdr}                     % para definir distintos tipos de cabeceras y pies de página

\usepackage{float}                             

\usepackage{amsmath}         % For advanced math typesetting
\usepackage{amsfonts}        % For additional mathematical fonts
\usepackage{amssymb}         % For more math symbols
\usepackage{amsthm}          % For theorem environments
\usepackage{empheq}          % For enhanced equation environments

\usepackage{color}           % For colored text
\usepackage{ragged2e}        % For ragged right alignment
\usepackage{titlesec}        % For customizing section title formats

\usepackage{tipa}            % For the thorn symbol
\usepackage[utf8]{inputenc} % For UTF-8 input encoding
\allowdisplaybreaks          % Allow page breaks within equations

\usepackage{enumitem}        % For customizable lists

\usepackage{tikz}  

\usepackage{hyperref}% For creating graphics
\usetikzlibrary{shapes.geometric,arrows.meta,arrows,positioning}  % Load specific TikZ libraries
\usetikzlibrary{tikzmark}   % For advanced TikZ marking

\titleformat{\paragraph}
  {\normalfont\normalsize\bfseries}{\theparagraph}{1em}{} % Format paragraph titles
\titlespacing*{\paragraph}
  {0pt}{3.25ex plus 1ex minus .2ex}{1.5ex plus .2ex} % Set spacing for paragraph titles

\begin{document}
\title{Ultracold Amplification Proposal for Parity Violation in Chiral Molecules}

\author{Daniel Martínez-Gil}
\thanks{Corresponding author}
\email{daniel.martinez@ua.es}
\affiliation{Departamento de F\'{\i}sica, Universidad de Alicante, Campus de San Vicente del Raspeig, E-03690 Alicante, Spain.}

\author{Pedro Bargueño}
\affiliation{Departamento de F\'{\i}sica, Universidad de Alicante, Campus de San Vicente del Raspeig, E-03690 Alicante, Spain.}

\author{Salvador Miret-Artés}
\affiliation{Instituto de Física Fundamental, Consejo Superior de Investigaciones Científicas, Serrano 123, 28006, Madrid, Spain}

\begin{abstract}
We propose a theoretical mechanism to indirectly detect the small parity-violating energy difference (PVED) between chiral enantiomers through a macroscopic enantiomeric excess observed in an ultracold gas. We consider that chiral molecules are formed resonantly via ultracold collisions of achiral diatomic molecules, with PVED inducing a slight asymmetry in the resonance energies of right- and left-handed configurations. After formation, chiral molecules evolve within a Bose-Einstein condensate (BEC), incorporating nonlinear interactions, tunneling between enantiomeric states, intrinsic PVED, and thermal conversion rates. These collective dynamics enable amplification of the microscopic bias into a global population imbalance. Using coupled rate equations, we show that, under realistic regimes, a complete enantiomeric excess can be achieved even for extremely small intrinsic asymmetries. We illustrate the model with concrete examples (HSOH, H$_2$Se$_2$, H$_2$Te$_2$), predicting observable enantiomeric excesses under plausible experimental conditions. We also consider non-PVED effects that could be amplified under the proposed mechanism, including electric and magnetic fields as well as thermal fluctuations, the latter being illustrated through the aforementioned molecular examples. Overall, our results suggest that ultracold physics could provide a new pathway to probe molecular parity violation, a fundamental weak effect that remains experimentally undetected.
\end{abstract}

\maketitle

\section{Introduction}

The existence of a parity-violating energy difference (PVED) between the two enantiomers of a chiral molecule constitutes one of the most intriguing manifestations of the electroweak interaction in molecular systems.  While parity violation has been unambiguously observed in atomic systems \cite{bouchibouchiatom}, its molecular counterpart remains experimentally undetected.  Theoretical studies predict that the PVED is extremely small for most molecules of chemical and biological relevance, lying below the energy resolution currently accessible in high-precision spectroscopy \cite{Quack2022}. 
Over the past decades, numerous experimental and theoretical efforts, approached from very different perspectives, have focused on the detection of the PVED in chiral molecules, ranging from high--resolution spectroscopy and precision measurements to detailed theoretical modelling of parity--violating effects in realistic molecular systems (see, for instance, Refs.~\cite{exp_fujiki1, exp_fujiki2, exp_fujiki3, exp_budker, exp_chardon, exp_MACDERMOTT200447, exp_MACDERMOTT200455, exp_schnell, exp_schwerd, ARIMONDO1977, Ziskind2002,Quack1986, Paris1999,Letokhov1975, Letokhov1976,PhysRevA.96.042119,Cournol, PhysRevX.11.031056, Lincoln,PhysRevX.13.041025,sahu2023detection,Dietiker,zurich1, zurich2, zurich3,experimentos,Landauetal,Eduardus,lee2024quantum, Bargueno2009, Gonzalo2010}). As a consequence, despite exhaustive experimental activity and the continuous development of increasingly sensitive spectroscopic techniques, no conclusive observation of a molecular PVED has yet been reported.

Beyond its fundamental interest, the PVED has long been discussed in connection with one of the most intriguing open questions at the interface between physics, chemistry, and biology: the origin of biomolecular homochirality.  Living systems display an almost complete dominance of one handedness, such as L--amino acids and D--sugars, and it has been proposed that an intrinsic parity-violating bias could have contributed to the initial symmetry breaking in prebiotic chemistry \cite{RevewHomochirality}.

A natural route toward enhancing the visibility of such electroweak effects could be provided by ultracold molecular physics  \cite{PerezRios2020}.  At very low temperatures, molecular collisions are governed by only a few well-defined quantum scattering channels and can be accurately controlled and characterized, while the long coherence times and the presence of narrow resonances make this regime particularly suitable for probing extremely small energy differences.
  In parallel, the rapid progress in the production, cooling, and trapping of molecules has opened the possibility of studying dense molecular systems \cite{jochimetal2003,krems2009coldmolecules}. In such systems, collective and nonlinear effects could convert microscopic energy differences into measurable population imbalances, thereby offering an attractive platform for the amplification of weak interactions \cite{avdeenkov2006resonance,flambaum2002molecular, Pedro2011PCCP,Pedro2012PRA}.

In this work, we propose an experimentally motivated mechanism in which the presence of the PVED can be inferred from a macroscopic observable, namely an enantiomeric excess that would vanish in the absence of parity violation. Our starting point is the formation of chiral molecules in ultracold $s$--wave collisions between two achiral diatomic molecules. Owing to the PVED, the formation probabilities of the left- and right-handed enantiomers are not exactly equal, although this difference remains extremely small at the microscopic level. As will be shown throughout this manuscript, this initial asymmetry alone is not enough to generate an experimentally observable enantiomeric excess (ee). It is therefore necessary to consider the subsequent post-scattering dynamics of the chiral molecules, which is assumed to evolve within a Bose-Einstein condensate (BEC). 

We incorporate the essential post-formation dynamics, including tunneling between enantiomers, the intrinsic PVED, and thermal effects, and we additionally introduce a mean-field nonlinear mechanism that naturally arises in a BEC. The interplay between these ingredients gives rise to an amplification mechanism through which a macroscopic and experimentally accessible enantiomeric excess can emerge from an extremely small parity-violating bias. Outside this collective and nonlinear regime, the same PVED would not lead to a detectable population imbalance. 

Remarkably, all simulations are performed using physically realistic values of the relevant parameters and on time scales compatible with the coherence times of ultracold BECs.

This manuscript is structured as follows: Section \ref{Sec1} introduces scattering dynamics, obtaining chiral molecules from diatomic ones. Section \ref{Sec2} details the post-scattering dynamics in a Bose-Einstein condensate. Section \ref{Sec3} shows the results of this manuscript using values of the parameters consistent with the literature, as well as a discussion of a possible amplification of non-PVED effects. Finally, Section \ref{Sec4} outlines the conclusions of the present work.

\section{Scattering dynamics}\label{Sec1}
We consider the resonant formation of a chiral molecular complex in the collision of two ultracold diatomic molecules. Since the formation of a chiral structure requires at least four atoms, the collision between two diatomic molecules represents the simplest elementary process leading to a chiral compound state.

In the ultracold regime, the collision dynamics is dominated by $s$-wave scattering. The cross section for the formation of a compound molecular resonance can be written in the Breit-Wigner \cite{FlambaumGinges2006} form as
\begin{equation}
\sigma(E-E_0)=\frac{\pi}{k^{2}}\,
\frac{\Gamma_{c}\Gamma}
{(E-E_{0})^{2}+\Gamma^{2}/4},
\end{equation}
where $k$ is the relative wave vector of the colliding molecules, $E$ is the collision energy, $E_{0}$ is the resonance energy in the absence of parity-violating effects, $\Gamma_{c}$ is the capture width, and $\Gamma$ is the total width of the resonance.

The parity-violating weak interaction in the chiral compound state leads to a small energy splitting between the left- and right-handed enantiomeric structures. Denoting the PVED by $\epsilon$, the resonance energies for the left- and right-handed configurations are shifted according to
\begin{equation}
E_{L}=E_{0}-\frac{\epsilon}{2}, \qquad
E_{R}=E_{0}+\frac{\epsilon}{2}.
\end{equation}
As a consequence, the cross sections for the formation of left- and right-handed molecules, $\sigma_{L}$ and $\sigma_{R}$, originating from achiral reactants, become slightly different, which can be quantified by introducing the asymmetry parameter \cite{FlambaumGinges2006}
\begin{equation}
P=\frac{\sigma_{R}-\sigma_{L}}{\sigma_{R}+\sigma_{L}}.
\end{equation}

Considering  the following Taylor expansion 
\begin{equation}
    \sigma(E-E_0\pm \frac{\epsilon}{2}) \approx \sigma(E-E_0)\pm \frac{\epsilon}{2}\frac{d \sigma(E-E_0)}{d (E-E_0)}+ \mathcal{O}(\epsilon^{2}),
\end{equation}
the maximum absolute value of the asymmetry parameter, reached at energies close to $E=E_{0}\pm\Gamma/2$, can be expressed as
\begin{equation}\label{pmax}
|P_{\mathrm{max}}|=\frac{2\epsilon}{\Gamma}.
\end{equation}

Note that Eq. \eqref{pmax} represents a difference in the probabilities for the formation of left- and right-handed chiral molecules in a single collision event. We therefore identify
\begin{equation}
|P_{\mathrm{max}}| = P_{L}-P_{R}.
\end{equation}

Furthermore, we assume that after the collision the only possible outcomes are the formation of either a left-handed or a right-handed chiral molecule. Consequently, the normalization condition
\begin{equation}
P_{L}+P_{R}=1
\end{equation}
must hold.
Under these assumptions, the probabilities for forming left- and right-handed molecules at the optimal collision energies $E=E_{0}\pm \Gamma/2$ are given by
\begin{align}
P_{L} &= \frac{1}{2}+\frac{\epsilon}{\Gamma},\\
P_{R} &= \frac{1}{2}-\frac{\epsilon}{\Gamma}.
\end{align}

We remark that we are working in the first-order approximation regime, so the condition $\epsilon \ll \Gamma$ must be fulfilled, maintaining $P_L,P_R>0$. This restriction holds for most molecular systems, as will be mentioned in the results.

\begin{figure}[ht]
\centering
\includegraphics[width= 15cm]{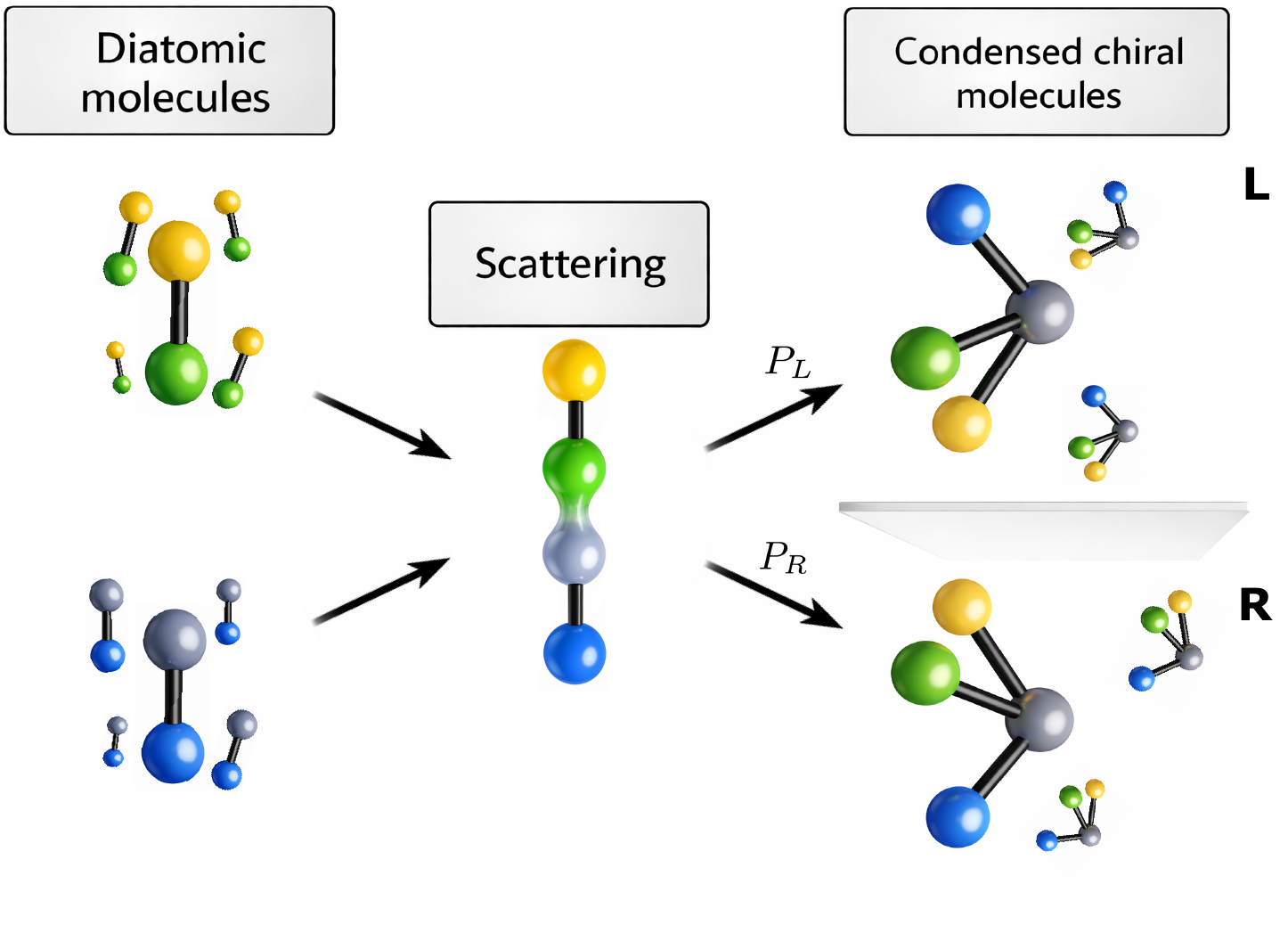}
\caption{\justifying Schematic diagram illustrating the scattering of two diatomic molecules leading to the formation of either the left- ($L$) or right-handed ($R$) enantiomer, with slightly different probabilities}
\label{esquema}
\end{figure}

Therefore, as schematically illustrated in Fig.~\eqref{esquema}, the basic idea of the present proposal is to study the collision between two achiral molecules at ultracold temperatures, such that resonant scattering leads to the formation of chiral molecules in either the left- or right-handed configuration, with slightly different formation probabilities. The formed chiral molecules are considered to evolve within a BEC, as will be explained in the following section.

Therefore, within this minimal description and neglecting any additional dynamical processes, we model the conversion of two colliding diatomic molecules into a chiral compound as an effective population transfer process governed by
\begin{align}
\frac{dN_T}{dt} &= -\Omega\, N_T , \\
\frac{dN_L}{dt} &= \Omega\, P_L\, N_T , \\
\frac{dN_R}{dt} &= \Omega\, P_R\, N_T ,
\end{align}
where $\Omega$ denotes an effective collision-induced conversion rate.

Here, $N_T$ represents the number of molecules available to be converted into chiral products, i.e., the total number of diatomic molecules participating in the collisions. In particular, since each scattering event involves two diatomic molecules, we write
\begin{equation}
N_T = 2N_{\mathrm{Di}},
\end{equation}
where $N_{\mathrm{Di}}$ is the number of remaining diatomic molecules in the gas. The quantities $N_L$ and $N_R$ denote the numbers of left- and right-handed chiral molecules produced, respectively.

Interestingly, this rate-equation model admits an analytical solution:
\begin{equation}
ee(t) = \frac{N_L(t)-N_R(t)}{N_L(t)+N_R(t)} = \frac{2\epsilon}{\Gamma},
\end{equation}
where $ee(t)$ is the enantiomeric excess, and it is equal to $\abs{P_{max}}$ for this case.

The enantiomeric excess, however, does depend on the resonance width $\Gamma$. This dependence is not favorable for detection purposes, since physically meaningful probabilities require $\Gamma$ to remain larger than twice the parity--violating energy shift, $\Gamma \geq 2\epsilon$. 

These considerations indicate that the formation process alone is not enough to generate a clearly detectable enantiomeric excess. It is therefore necessary to include the post-scattering dynamics of the chiral molecules and to investigate whether additional physical ingredients can provide an efficient amplification mechanism capable of enhancing the enantiomeric excess to experimentally observable levels.

\section{Post-scattering dynamics in a Bose-Einstein condensate}\label{Sec2}

Let us assume that the produced molecular sample evolves within a Bose--Einstein condensate, which is a crucial ingredient of our model, since the internal dynamics of the molecular ensemble is no longer strictly governed by a linear Schr\"odinger equation. Instead, mean-field interactions between molecules give rise to an effective nonlinear dynamics governed by a Gross-Pitaevskii-type equation \cite{vardi}, as we will describe in this section.

We first introduce the relevant variables associated with the internal dynamics of chiral molecules. In this work, the internal chiral degree of freedom is modeled as a two-level system (TLS) described by the Hamiltonian
\begin{equation}\label{hamilsigmas}
    \hat{H} = \delta \hat{\sigma}_x + \epsilon \hat{\sigma}_z,
\end{equation}
which is commonly employed in the description of chiral molecules \cite{Harrisystodolsky}.

The parameter $\delta$ is associated with tunnelling between the left and right configurations and is therefore inversely related to the tunnelling time. The parameter $\epsilon$ quantifies the energetic asymmetry between the two localized configurations.

Within this framework, the general state of the TLS, written in the chiral basis, can be expressed as
\begin{equation}
    \ket{\psi(t)} = a_L(t)\ket{L} + a_R(t)\ket{R},
\end{equation}
leading to the following set of coupled time-dependent Schr\"odinger equations,
\begin{align}
    i\hbar \dot{a}_L &= \epsilon a_L + \delta a_R, \\
    i\hbar \dot{a}_R &= \delta a_L - \epsilon a_R .
\end{align}

Therefore, the parameters $\epsilon$ and $\delta$ constitute two fundamental quantities governing the post-scattering internal dynamics of the chiral molecules.

As we have previously mentioned, working in a BEC implies effective nonlinear dynamics. It has been shown \cite{daniDissipativeTLS} that the corresponding self-interaction term can be incorporated into the above two-level description as an effective nonlinear contribution to the equations of motion, 
\begin{eqnarray}
i\hbar \dot{a}_L &=& \epsilon a_L + \delta a_R 
+ \frac{1}{2}\, a_L\, \Lambda \left(|a_L|^2-|a_R|^2\right), \label{aL}\\
i\hbar \dot{a}_R &=& \delta a_L - \epsilon a_R 
- \frac{1}{2}\, a_R\, \Lambda \left(|a_L|^2-|a_R|^2\right), \label{aR}
\end{eqnarray}
where the parameter $\Lambda$ characterizes the strength of the nonlinear mean-field interaction.

The nonlinear term proportional to $\Lambda$ acts similarly to the asymmetry parameter $\epsilon$. It is therefore convenient to introduce an effective PVED,
\begin{equation}\label{epsiloneff}
    \epsilon_{\mathrm{eff}} = \epsilon + \frac{1}{2}\Lambda\left(|a_L|^2-|a_R|^2\right),
\end{equation}
which allows the Gross-Pitaevskii dynamics to be recast in a form formally identical to the linear two-level Schr\"odinger equations,
\begin{align}
    i\hbar \dot{a}_L &= \epsilon_{\mathrm{eff}} a_L + \delta a_R, \\
    i\hbar \dot{a}_R &= \delta a_L - \epsilon_{\mathrm{eff}} a_R .
\end{align}

We are now in a position to introduce a more complete model that incorporates post-scattering dynamics. The population dynamics is described by the coupled rate equations 
\begin{align}
    \frac{d N_T}{dt} &= -\Omega N_T ,\\
    \frac{d N_L}{dt} &= \Omega P_L N_T - K_{LR} N_L + K_{RL} N_R ,\\
    \frac{d N_R}{dt} &= \Omega P_R N_T + K_{LR} N_L - K_{RL} N_R ,
\end{align}
where the rate constants $K_{LR}$ and $K_{RL}$ describe the interconversion processes from $L$ to $R$ and from $R$ to $L$, and the populations are defined as $N_L = \abs{a_L}^2$, $N_R = \abs{a_R}^2$. Again, we emphasize that the post-scattering dynamics takes place inside a Bose-Einstein condensate. As a consequence, the rate constants must consider physical effects intrinsic to a BEC, as will be discussed in the following.

In the absence of parity-violating effects, these rates would be symmetric and directly related to the tunnelling parameter $\delta$. However, parity violation introduces an energetic bias between the two chiral configurations, leading to asymmetric transition rates. As discussed by Kondepudi and Nelson in the context of biomolecular chirality, the rate constants can be exponentially modified by the PVED \cite{KondepudiNelson1985}. Therefore, in the present model, we consider these rates as
\begin{align}
    K_{LR} &= \delta \exp\left(\frac{\epsilon_{\mathrm{eff}}}{k_B T}\right), \label{KLR2}\\
    K_{RL} &= \delta \exp\left(-\frac{\epsilon_{\mathrm{eff}}}{k_B T}\right). \label{KRL2}
\end{align}

The resulting set of coupled differential equations therefore incorporates scattering processes, tunnelling dynamics, parity-violating energy difference, thermal effects, and nonlinear mean-field interactions arising from the BEC, within a simple model.

\section{Results}\label{Sec3}
We first remark that, although BECs of chiral molecules have not yet been realized, this may become feasible in the near future, as condensates have already been successfully produced, for instance, with diatomic molecules \cite{Bigagli2024}. Therefore, in the following, we present results based on parameter values reported in the literature, with the aim of enabling a possible future experimental realization of our predictions.

We begin by introducing the effective collision-induced conversion rate $\Omega$. This rate can be expressed as
\begin{equation}
    \Omega = k \, n,
    \label{eq:omega}
\end{equation}
where $k$ is the two-body reaction (or inelastic loss) rate constant (in units of cm$^3$ s$^{-1}$), and $n$ is the number density of the colliding molecules (in cm$^{-3}$).
Experimental studies of ultracold molecular collisions report typical values of $k$ in the range $\approx (10^{-10} - 10^{-12})\,$cm$^3\,$s$^{-1}$ \cite{Rate_constants1,Rate_constants3,Rate_constants4,Rate_constants2}. At the same time, the number of densities achieved in ultracold molecular samples and molecular BECs typically lie between $\approx (10^{10} - 10^{12})\,$cm$^{-3}$ \cite{Rate_constants2,densidades1,densidades2}.

Combining these values, we obtain an estimated conversion rate of 
$\Omega \approx 10^{-2}-10^{2}\,\mathrm{Hz}$ under typical experimental conditions. 
This range is consistent with the characteristic timescales reported in ultracold
reactive--collision experiments and therefore provides a physically motivated
choice for $\Omega$ in our numerical simulations. 

In the present work, we fix the effective conversion rate to the representative value $\Omega = 100\,\mathrm{Hz}$, corresponding to a high-density regime. This choice is motivated by our interest in fast conversion dynamics, which facilitate the generation of a measurable enantiomeric excess within experimentally accessible time scales. In particular, previous studies have shown that coherence times on the order of a few seconds, and typically around $3\,\mathrm{s}$, can be achieved in optimized ultracold atomic and molecular experiments (see, e.g., \cite{time1,time2}). In order to remain within this experimentally realistic coherence window, we therefore restrict our numerical simulations to 3 s throughout this work.

Furthermore, the literature reports a wide range of molecular candidates for which the relevant $\epsilon$ and $\delta$ parameters have been estimated (see \cite{quack2008}). In terms of frequency, $\epsilon$ is estimated to range from about $10^{-4}$ to $90$ Hz, while the tunneling parameter $\delta$ spans a much wider interval. Experimentally, most accessible $\delta$ values vary roughly from $3\cdot 10^{-15}$ to $3\cdot 10^{11}$ Hz, depending on the specific molecular system. In addition, estimates for the non-linear parameter $\Lambda$ are also available and are typically found to depend on $\delta$. For instance, values of $\Lambda \leq
 30 \,\delta$ are commonly assumed in the literature \cite{pendulo1,Pendulo2,Pendulo3,Oberthaler2005}).

\begin{figure}[H]
	\begin{subfigure}[b]{0.49\textwidth}
		\includegraphics[width=\textwidth, height=7cm]{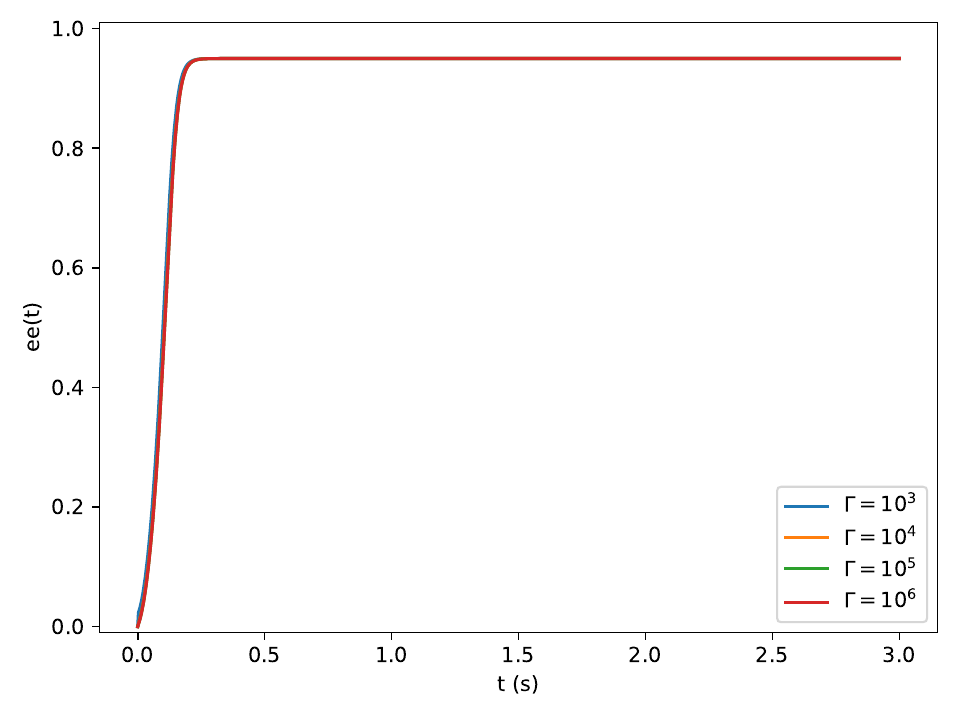}
	
	\end{subfigure}
	\hfill
	\begin{subfigure}[b]{0.49\textwidth}
		\includegraphics[width=\textwidth, height=7cm]{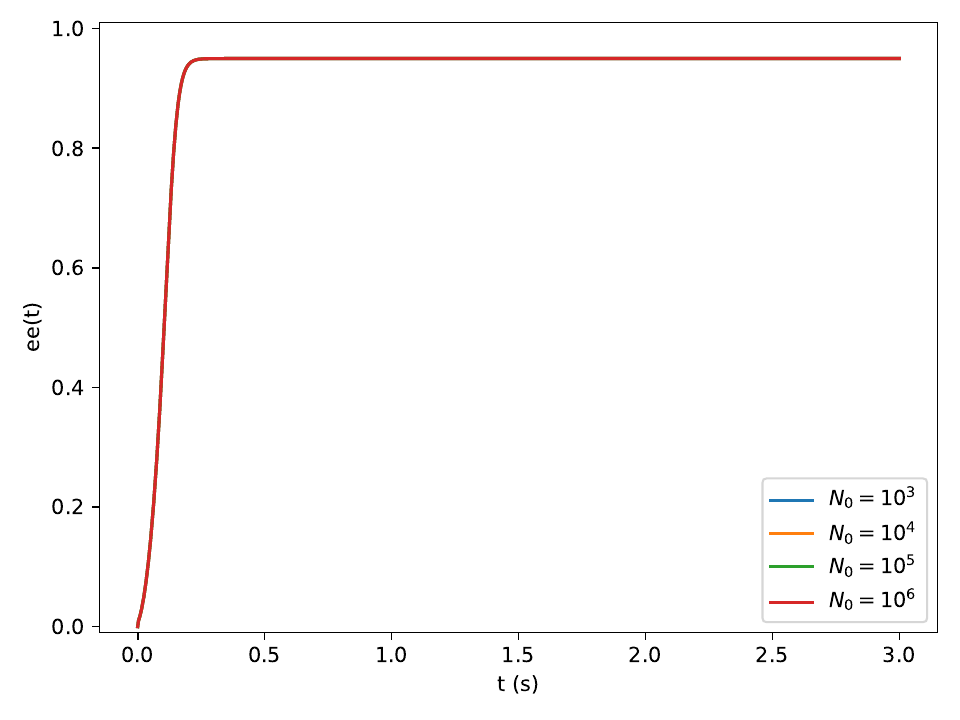}

	\end{subfigure}
	\caption{\justifying Time variation of enantiomeric excess for different values of $\Gamma$ (left panel) and $N_0$ (right panel). Simulation parameters: $\epsilon = 10$ Hz, $\delta = 10$ Hz, $\Lambda= 15\delta$.}\label{variandogamma}
\end{figure}

The value of the resonance width must ultimately be determined experimentally for a given molecular system. As a representative reference, very narrow widths on the order of a few kilohertz have been reported in ultracold atomic systems. In particular, in $^{133}$Cs, $s$-wave resonances with widths as small as $\Gamma \simeq 3.5\,\mathrm{kHz}$ have been observed~\cite{Herbig2003}.
In the following, we therefore adopt this value as a reference width in our numerical simulations. We emphasize that, although the resonance width may strongly depend on the specific molecular species and on the particular experimental conditions, and will in general be larger than this narrow reference resonance, our results remain qualitatively unchanged over a wide range of $\Gamma$, as illustrated in the left panel of Fig.~\eqref{variandogamma}.

Finally, we emphasize that, as within the previous scattering model, the enantiomeric excess is independent of the total number of initially available diatomic molecules, $N_0$, participating in the condensate (see right panel of Fig. \eqref{variandogamma}). Consequently, the value of $N_{0}$ does not affect the time evolution of the enantiomeric excess, provided that a BEC can be formed. We therefore consider an initial molecular population $N_{0}$ large enough to support the formation of a BEC, without imposing any further constraint on its precise value. In addition, we consider a temperature of the order of $T \simeq 1\,\mathrm{nK}$, consistent with current ultracold molecular and atomic experiments. 
This choice is motivated by both experimental feasibility and by the role played by temperature in our model. On the one hand, temperatures in the nanokelvin regime are routinely achieved in present-day BEC experiments, and temperatures in the picokelvin range have already been demonstrated under specially optimized conditions~\cite{picokelvin1,picokelvin2}. On the other hand, as can be seen from Eqs.~\eqref{KLR} and~\eqref{KRL}, the temperature directly controls the sensitivity of the rate constants to the effective bias $\epsilon_{\mathrm{eff}}$, and therefore acts as an amplification factor for the resulting enantiomeric excess. Lower temperatures lead to a stronger imbalance between the two rates and consequently to a more pronounced enantiomeric excess. While temperatures of the order of $10\,\mathrm{nK}$ would be experimentally more accessible, they would produce a weaker and less visible effect within the present model. Conversely, operating in the picokelvin regime would further enhance the signal, but at the cost of significantly increased experimental complexity. We therefore adopt $T\simeq1\,\mathrm{nK}$ as a realistic balance between experimental accessibility and amplification efficiency. All the parameters, both with the literature values and the ones used in this work, are summarized in Table \eqref{tab:data_exp}.

\begin{table}[H]
\centering
\begin{tabular}{ccc}
\hline
Parameter & Value & This work\\
\hline
\noalign{\vskip 1mm}
$\Omega$   &$\approx (10^{-2}-10^{2})\,\mathrm{Hz}$ \cite{Rate_constants1,Rate_constants3,Rate_constants4,Rate_constants2}   & $100\,\mathrm{Hz}$ \\
$\epsilon$   & ($10^{-4} -$ $90)$ Hz \cite{quack2008} & $(10^{-4} -$ $90)$ Hz  \\
$\delta$   & $ \approx (3\cdot 10^{-15} -$ $3\cdot 10^{11})$ Hz \cite{quack2008} &  \quad$ (10^{-1} - 10^{7})$ Hz \\
$\Lambda$   & $ \leq
 30 \,\delta$  \cite{pendulo1,Pendulo2,Pendulo3,Oberthaler2005} &  $ (0 - 30) \delta$ \\
$\Gamma$   & $\gtrsim3500$ Hz \cite{Herbig2003}  &   $3500$ Hz  \\
$N_0$   & $<10^7$  & $10^6$  \\
$T$   & $\geq
38$ pK \cite{picokelvin1,picokelvin2} & $1$ nK  \\

\hline
\end{tabular}
\caption{Summary of the parameter values used in this work, together with representative values reported in the literature. See text for details.}\label{tab:data_exp}
\end{table}

First of all, we will attempt to quantitatively evaluate how the enantiomeric excess
varies as a function of each of the model's variable parameters.

\begin{figure}[H]
	\begin{subfigure}[b]{0.49\textwidth}
		\includegraphics[width=\textwidth, height=7cm]{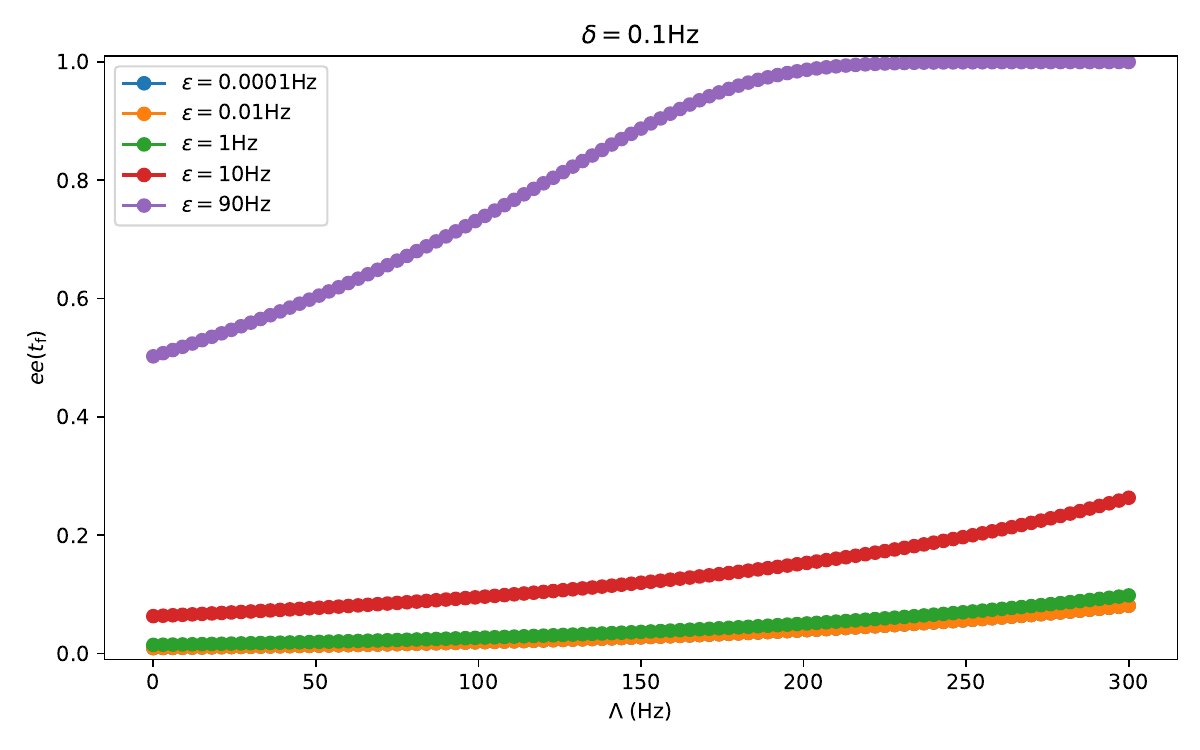}
	
	\end{subfigure}
	\hfill
	\begin{subfigure}[b]{0.49\textwidth}
		\includegraphics[width=\textwidth, height=7cm]{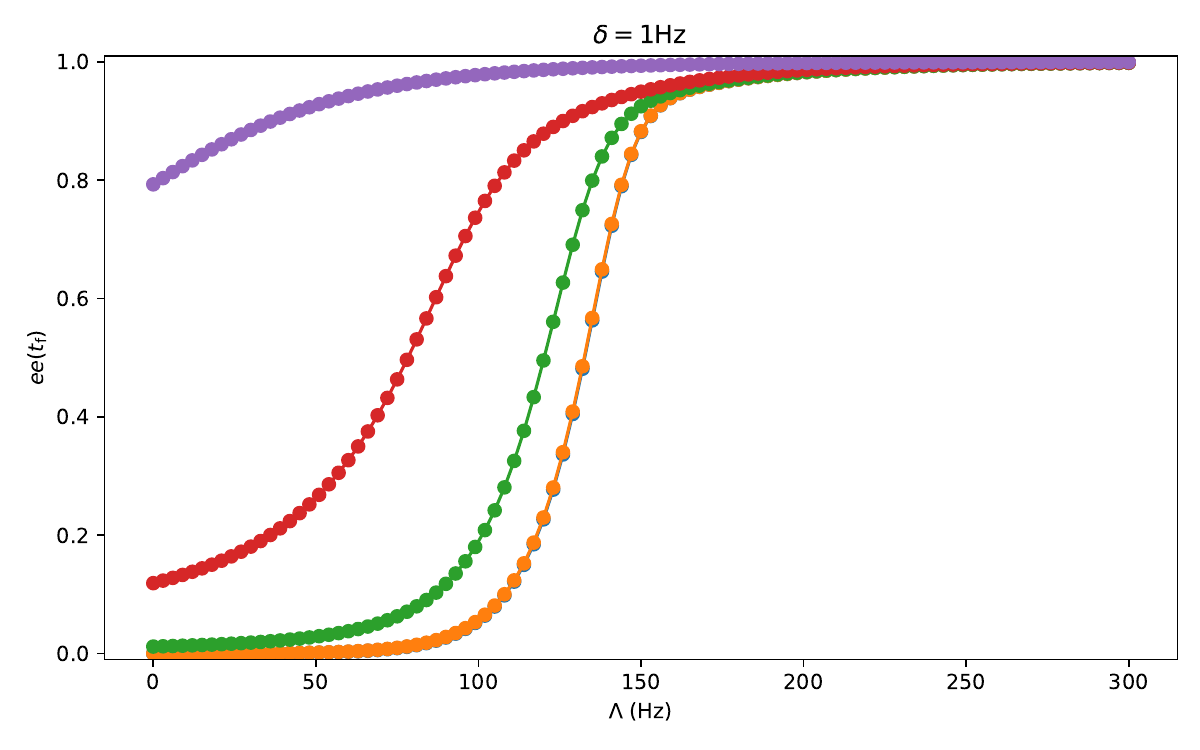}

	\end{subfigure}
    \begin{subfigure}[b]{0.49\textwidth}
		\includegraphics[width=\textwidth, height=7cm]{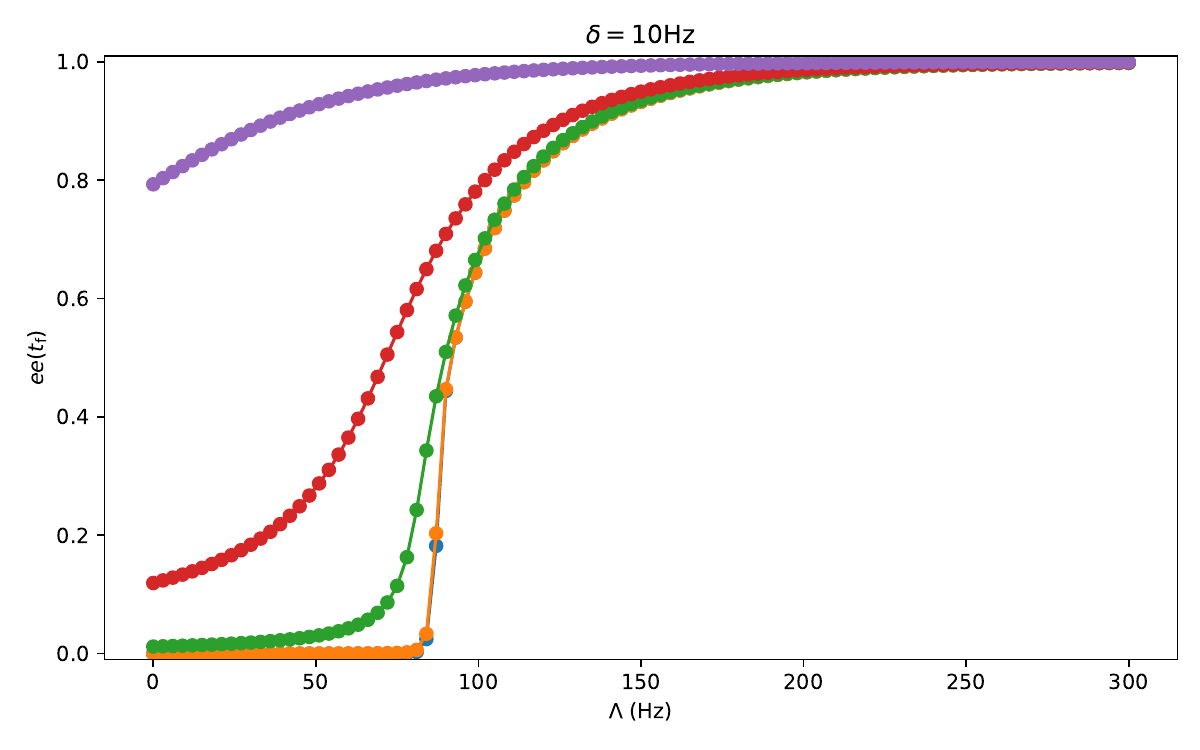}
	
	\end{subfigure}
	\hfill
    \begin{subfigure}[b]{0.49\textwidth}
		\includegraphics[width=\textwidth, height=7cm]{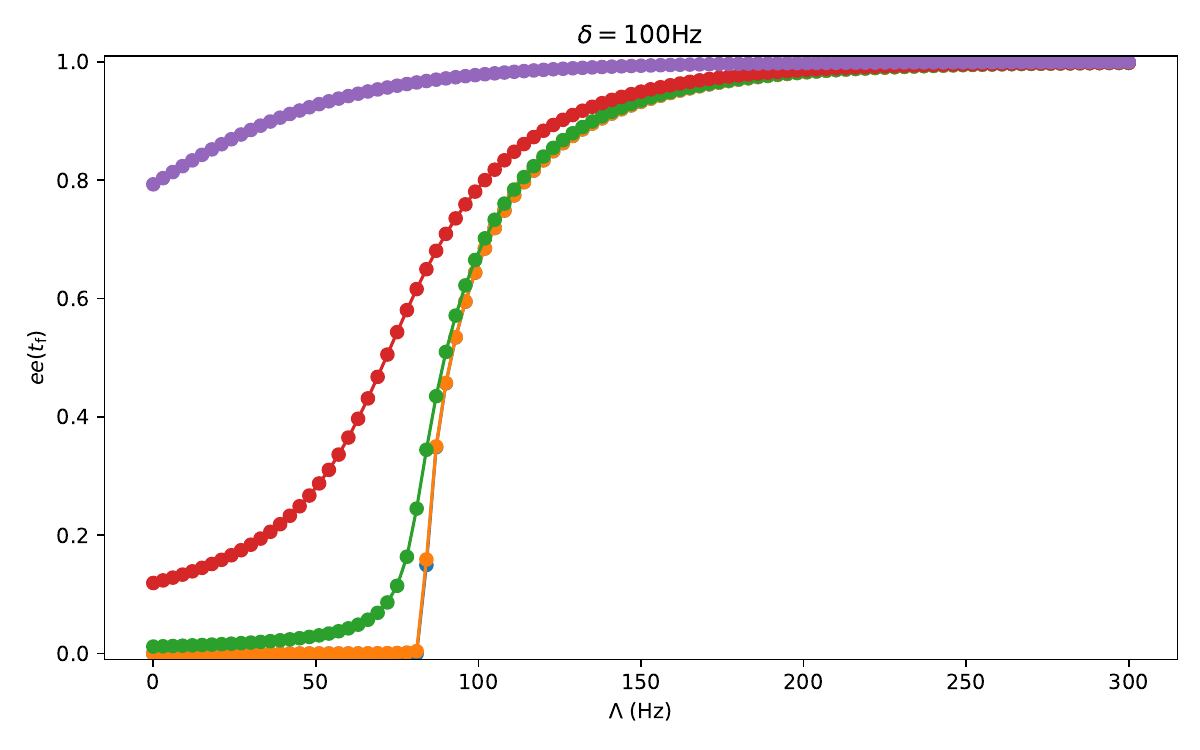}
	
	\end{subfigure}
	\caption{Enantiomeric excess at $t_f = 3$ s, considering different values for $\epsilon$, $\delta$, and $\Lambda$. $\Gamma = 3500$ Hz.}\label{variandodeltaepsilonlambda}
\end{figure}

Fig. (\ref{variandodeltaepsilonlambda}) shows how the enantiomeric excess at a final (and characteristic coherence) time $t_f$ depends on the model parameters $\delta$, $\epsilon$, and $\Lambda$. In these simulations, we sample representative values of $\delta$ in the interval $[0.1,100]\,$Hz, $\Lambda$ in the interval $[0,300]\,$Hz, and vary $\epsilon$ over ranges motivated by the literature (see text below).

Several clear conclusions emerge from the results. First, when the nonlinear mean-field interaction is absent ($\Lambda=0$ Hz), the final enantiomeric excess depends only on the PVED. Smaller values of $\epsilon$ produce vanishing enantiomer excess, while larger values of $\epsilon$ yield a non-zero excess. In our parameter set, the excess ranges from essentially zero for the smallest $\epsilon$ up to approximately $0.35$ for the largest $\epsilon$.
Physically, larger values of $\Lambda$ increase the effective interconversion rates and, through the population-dependent $\epsilon_{eff}$ parameter (Eq. \eqref{epsiloneff}), reinforce any small initial bias until one enantiomer dominates. It is worth noting that this enantioselection effect is similar to the well-known self-trapping effect, whereby increasing $\Lambda$ causes one enantiomer to become localized \cite{pendulo1, Pendulo2,Pendulo3, daniDissipativeTLS}.
Second, we also note that increasing the tunnelling parameter $\delta$ reduces the value of the resonance width $\Gamma$ required to achieve a regime of complete enantiomeric excess. In particular, the curve corresponding to $\delta = 10$ Hz represents the realistic interval discussed above, $\Lambda \leq 30\delta$, and already demonstrates that a fully developed enantiomeric excess can be obtained under experimentally plausible conditions. For larger values of $\delta$, the resulting enantiomeric excess curves are found to be close to the complete enantioselection, indicating that further increases in the tunnelling rate do not lead to substantial changes. Therefore, $\delta \gtrsim 10\,\mathrm{Hz}$ is enough for the observation of the predicted effects.
Finally, we note an additional property observed in the simulations: the curve corresponding to $\epsilon=10^{-4}$ Hz (blue) is essentially overlapped with that for $\epsilon=10^{-2}$ Hz (orange) across the considered parameter space. Within our numerical resolution, these two lines are practically indistinguishable, which indicates that, for sufficiently small $\epsilon$, the nonlinear amplification mechanism leads to a nearly complete enantiomeric excess provided that $\delta$ is sufficiently large, independently of further reductions of the PVED.

Therefore, as is shown in Fig. \eqref{variandodeltaepsilonlambda}, for any value of $\epsilon$ within the range reported in the literature, an experimentally observable enantiomeric excess can be obtained provided that the tunnelling parameter of the formed chiral molecule satisfies (preferably) $\delta \gtrsim 10\,\mathrm{Hz}$. According to the data reported in Ref.~\cite{quack2008}, at least about one half of known chiral molecules fulfil this condition. We therefore consider, as representative examples, a set of molecules whose tunnelling parameters satisfy $\delta \gtrsim 10\,\mathrm{Hz}$. The molecular species considered in the following are listed in Table 
\ref{tab:real_molecules}.

\begin{table}[H]
\centering
\begin{tabular}{lcc}
\hline
Molecule & $\epsilon$ (Hz) & $\delta$ (Hz) \\
\hline
\noalign{\vskip 1mm}
HSOH   & $1.2\times10^{-2}$ & $6\times10^{7}$ \\
H$_2$Se$_2$ & $6$                 & $3\times10^{4}$ \\
H$_2$Te$_2$ & $90$                & $9\times10^{2}$ \\
\hline
\end{tabular}
\caption{Representative chiral molecules considered in this work and corresponding PVED  and tunnelling parameters expressed in Hz. Data from \cite{quack2008}.}\label{tab:real_molecules}
\end{table}

\begin{figure}[H]
\centering

% ---------- fila 1 ----------
\begin{subfigure}[b]{0.32\textwidth}
    \centering
    \includegraphics[width=\textwidth]{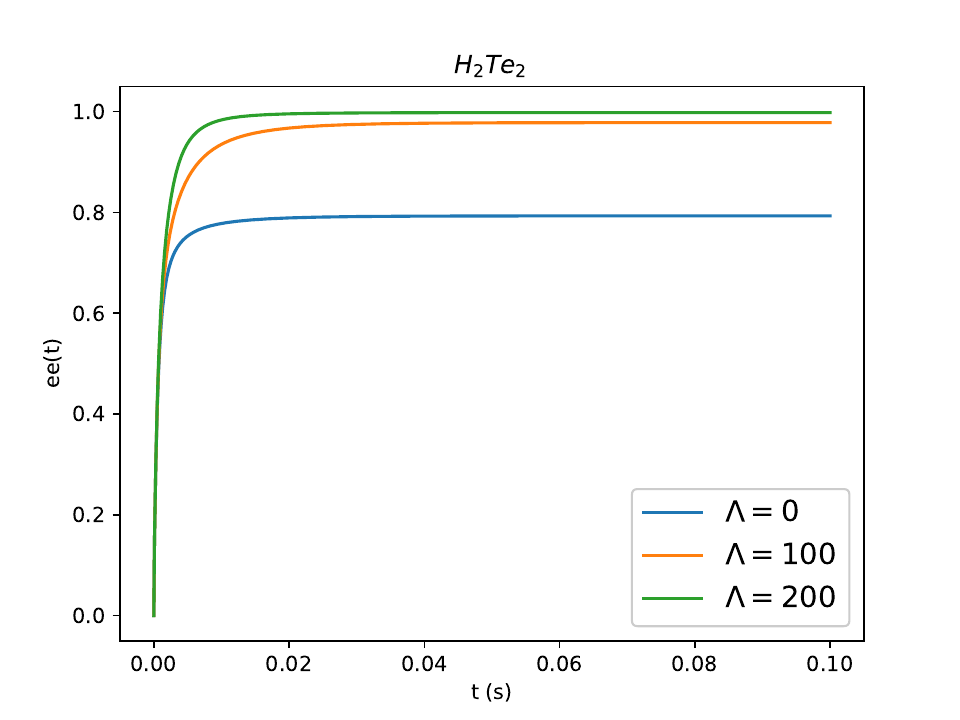}
\end{subfigure}
\hfill
\begin{subfigure}[b]{0.32\textwidth}
    \centering
    \includegraphics[width=\textwidth]{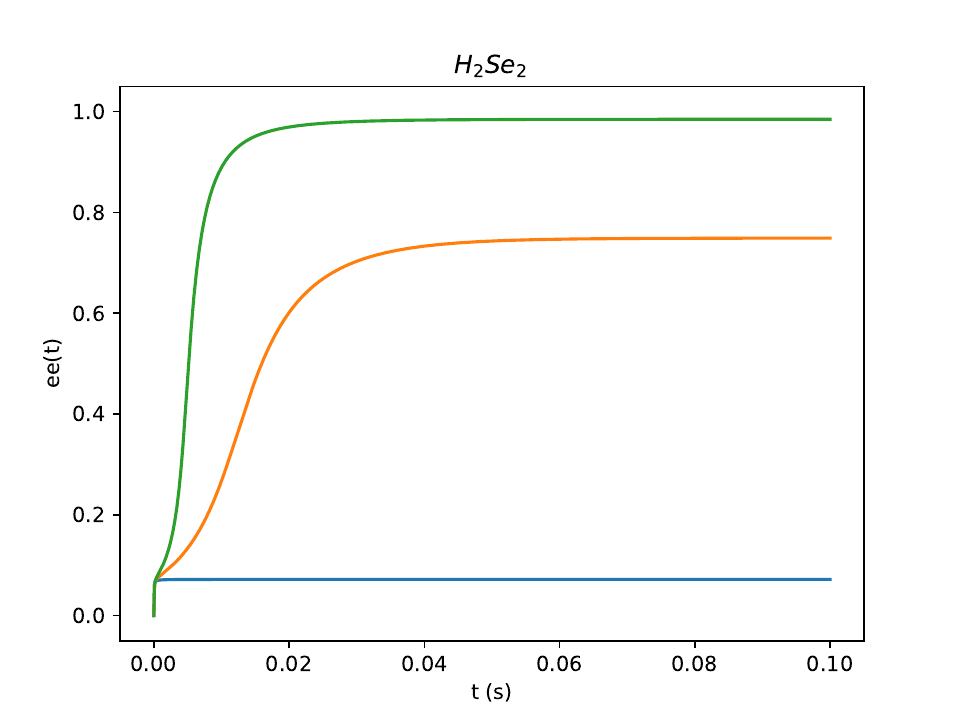}
\end{subfigure}
\hfill
\begin{subfigure}[b]{0.32\textwidth}
    \centering
    \includegraphics[width=\textwidth]{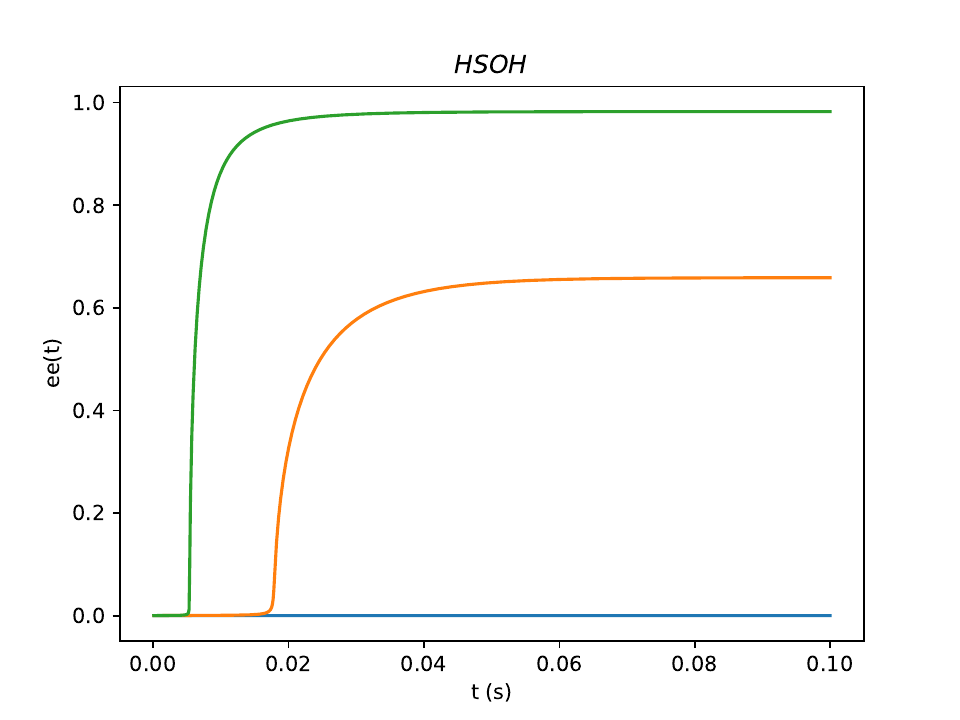}
\end{subfigure}

\vspace{3mm}

% ---------- fila 2 ----------
\begin{subfigure}[b]{0.32\textwidth}
    \centering
    \includegraphics[width=\textwidth]{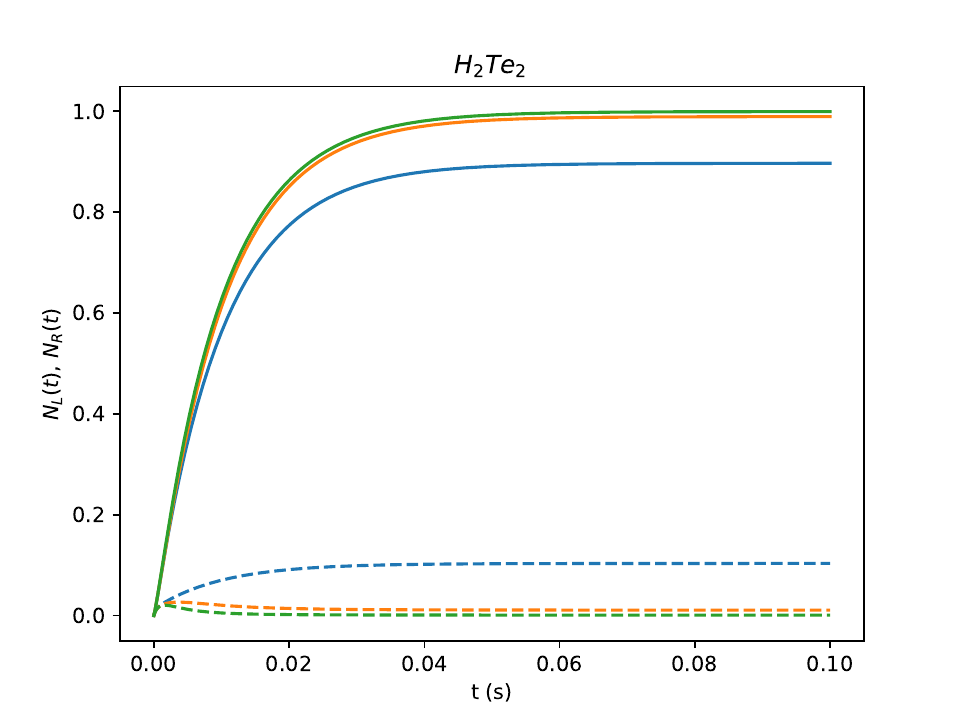}
\end{subfigure}
\hfill
\begin{subfigure}[b]{0.32\textwidth}
    \centering
    \includegraphics[width=\textwidth]{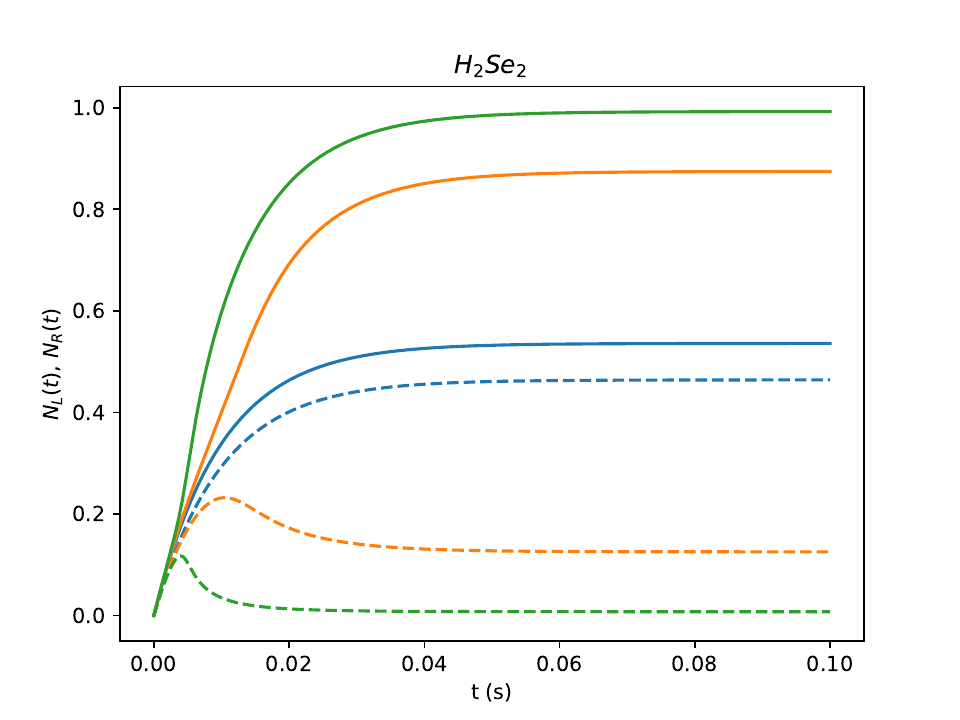}
\end{subfigure}
\hfill
\begin{subfigure}[b]{0.32\textwidth}
    \centering
    \includegraphics[width=\textwidth]{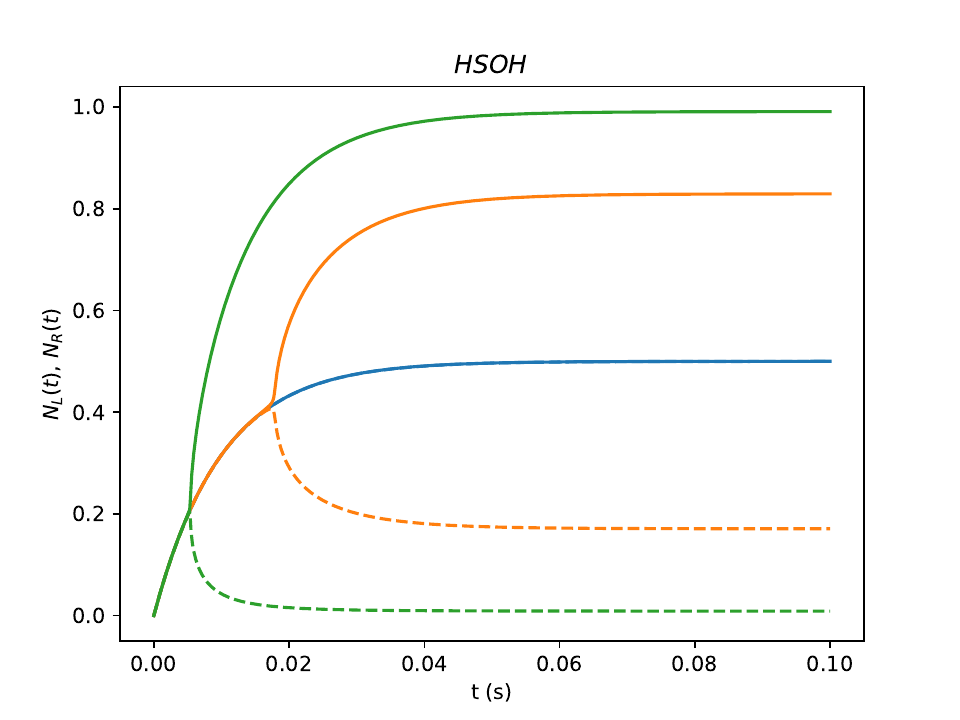}
\end{subfigure}

\caption{The results are shown for three different chiral molecules, ordered from left to right as H$_2$Te$_2$, H$_2$Se$_2$, and HSOH. The first row displays the enantiomeric excess, while the second row shows the time evolution of the populations of the two enantiomers, $N_L$ and $N_R$. As indicated in the legend, the blue curves correspond to $\Lambda = 0$, the orange curves to $\Lambda = 100$ Hz, and the green curves to $\Lambda = 200$ Hz. In the second row, solid lines represent the population of the left-handed enantiomer ($N_L$), whereas dashed lines represent the population of the right-handed enantiomer ($N_R$). The values of $\epsilon$ and $\delta$ used for each molecule are reported in Table~\eqref{tab:real_molecules}. In all cases, $\Gamma = 3500\,\mathrm{Hz}$ is considered.}
\label{molreales}
\end{figure}

In Fig.~\eqref{molreales} we show both the enantiomeric excess and the time evolution of the populations $N_L$ and $N_R$ for these chiral molecules, considering the three cases $\Lambda = 0, 100, 200$ Hz. We emphasize that, in all simulations, the condition $\Lambda < 30\,\delta$ is fulfilled. The figure illustrates that, for these specific molecular systems, a complete enantiomeric excess can be achieved for the three molecules when a sufficiently large nonlinear coupling is considered, for instance, for $\Lambda = 200$ Hz. In contrast, in the absence of nonlinear interactions ($\Lambda = 0$ Hz), the final enantiomeric excess is determined by the value of $\epsilon$ (we have to remind the reader that we are considering a resonance width of $\Gamma = 3500$ Hz). As a consequence, molecules with large PVED, such as H$_2$Te$_2$, can reach enantiomeric excess values of the order of $75\%$, whereas molecules with much smaller $\epsilon$, such as HSOH, remain essentially racemic in the absence of $\Lambda$. It is also remarkable that the enantiomeric excess exhibits a sudden increase after an initial period of smooth evolution. This behavior reflects the nonlinear character of the model, which can lead to a qualitative change in the system dynamics.

It is important to emphasize that, since all the simulations presented in this work are performed on time scales compatible with the coherence times of a BEC, decoherence effects are not expected to play a relevant role in the considered time-scales.
If one were instead interested in describing the long-time dynamics beyond the maximum simulation times considered here (of the order of three seconds), decoherence could be incorporated phenomenologically by introducing a dissipative or friction-like term acting on the population difference, $N_L-N_R$ (see, e.g., \cite{penate2013langevin}). In practice, this can be effectively modeled as a gradual suppression of the tunneling parameter $\delta$, which reduces the coherent exchange between enantiomers over time. However, such a term would have little impact on the populations once they have reached a complete enantiomeric excess, since at that stage, tunneling dynamics no longer play a significant role.

Nevertheless,  if the study is focused on explicitly tracking quantum coherences between the left- and right-handed states, decoherence would primarily manifest as phase diffusion. In this case, the global relative phase of the molecular ensemble would evolve and fluctuate over time, producing a gradual loss of coherence (dephasing) \cite{Oberthaler2005, ruostekoski1999}.

\subsection{Non-PVED amplification}

\begin{figure}[H]
\centering
% ---------- fila 1 ----------
\begin{subfigure}[b]{0.32\textwidth}
    \centering
\includegraphics[width=\textwidth]{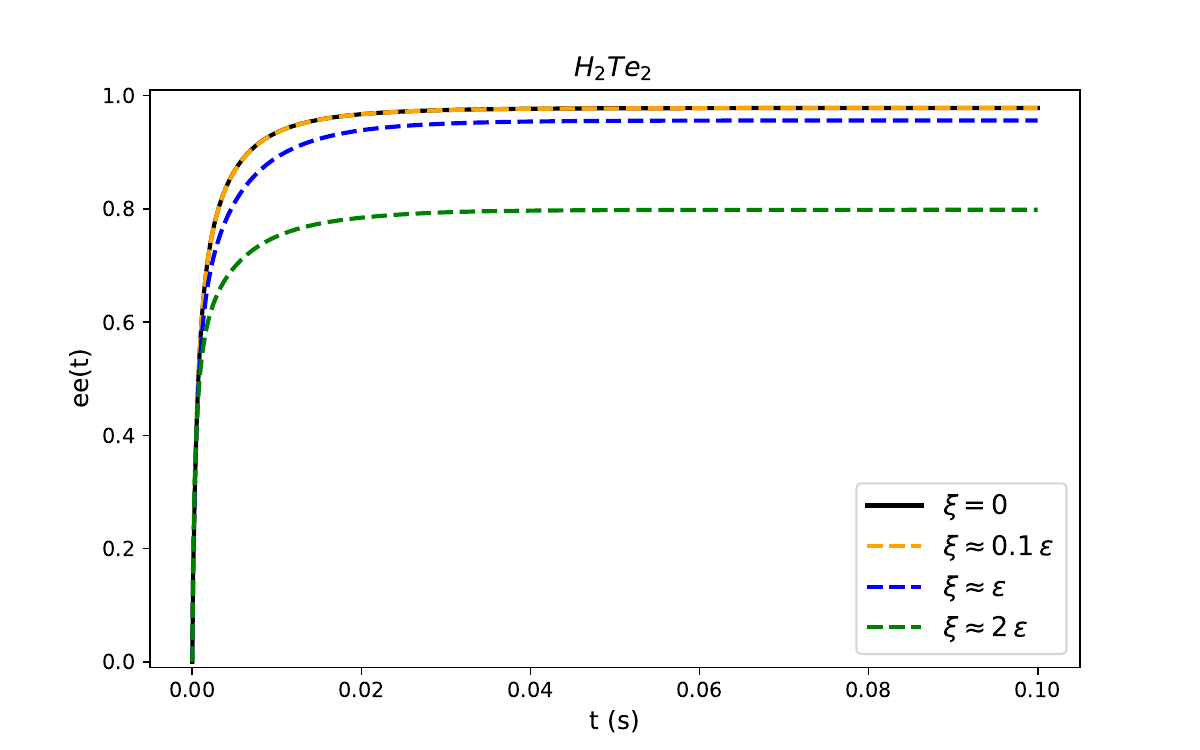}
\end{subfigure}
\hfill
\begin{subfigure}[b]{0.32\textwidth}
    \centering
    \includegraphics[width=\textwidth]{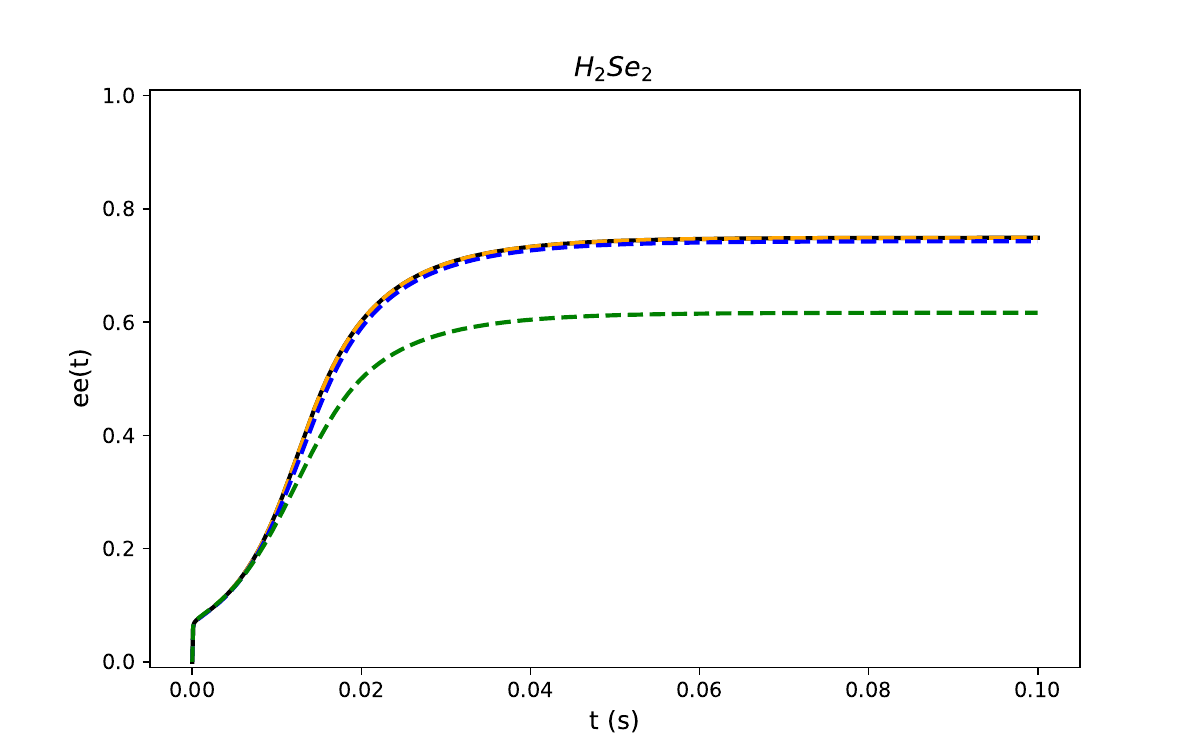}
\end{subfigure}
\hfill
\begin{subfigure}[b]{0.32\textwidth}
    \centering
\includegraphics[width=\textwidth]{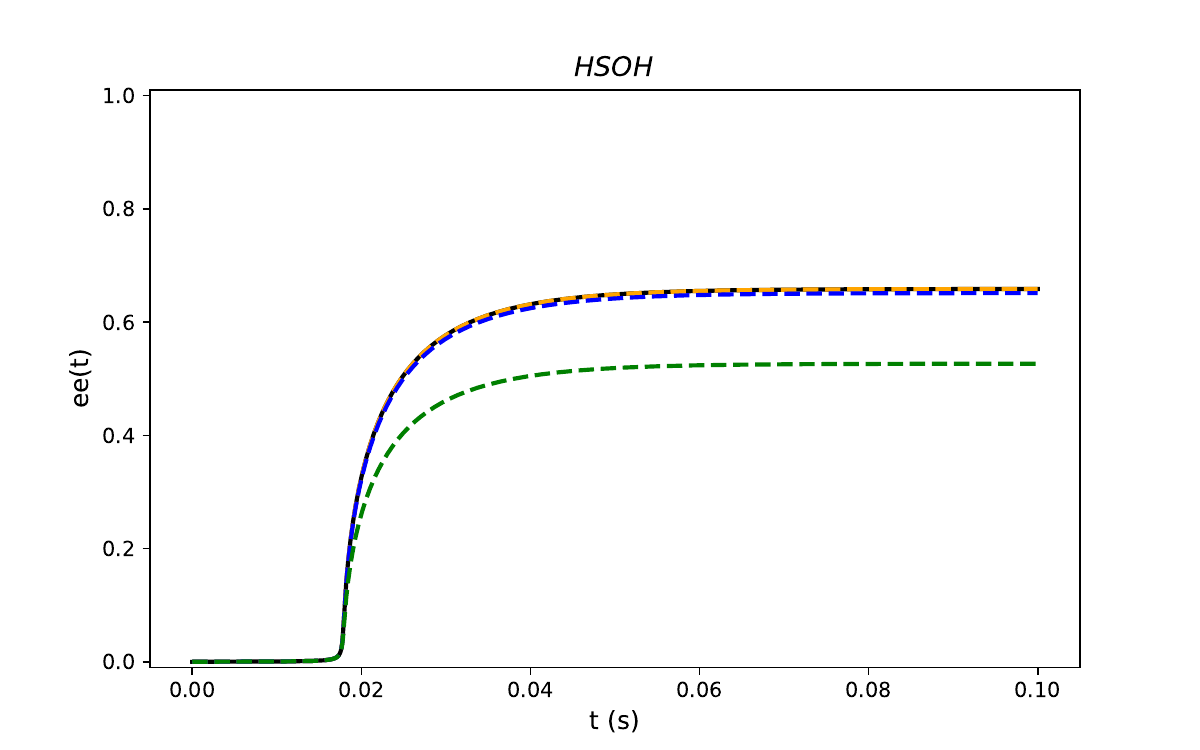}
\end{subfigure}
\caption{These results show the dependence of the enantiomeric excess considering a stochastic noise, which is amplified by the considered mechanism. We consider three different chiral molecules as examples, varying the fluctuations $\xi$ from 0 to $\approx 2 \epsilon$. In all results, we consider $\Lambda = 100$ and we average 2000 times over independent noise realizations.}
\label{fluctuaciones}
\end{figure}

We now consider effects of random fluctuations, for instance those arising from thermal noise, and their influence on the generation of an enantiomeric excess. This question naturally emerges from the nonlinear character of the model, since mechanisms different from the PVED may also be amplified by the nonlinear dynamics. In particular, thermal fluctuations could induce a chiral bias that competes with the parity-violating contribution.
\\
\\
It is usual in these models to consider thermal fluctuations as stochastic noise affecting the reaction rates (see Refs.~\cite{Hochberg2022PVED,Hochberg2023Noise}, where the effects of noise and PVED are directly compared). Following this approach, we introduce noise in the temperature-dependent term of the reaction rates. Specifically, we consider
\begin{align}
K_{LR} &= \delta \exp\left(\frac{\epsilon_{\mathrm{eff}}+\xi}{k_B T}\right), \label{KLR}\\
K_{RL} &= \delta \exp\left(-\frac{\epsilon_{\mathrm{eff}}+\xi}{k_B T}\right), \label{KRL}
\end{align}
where $\xi$ denotes a stochastic noise variable. We model $\xi$ as Gaussian white noise with zero mean. Three different noise amplitudes are considered: $\xi \approx 0.1\epsilon$, $\xi \approx \epsilon$, and $\xi \approx 2\epsilon$. Here, the symbol ``$\approx$'' indicates that the Gaussian distribution is chosen such that there is a $99\%$ probability that the noise magnitude lies within the corresponding interval.
\\
\\
To reduce the statistical uncertainty associated with individual realizations, we average over a large number of independent noise realizations. In particular, all results presented below correspond to averages over $2000$ realizations. It is important to note that the effect of the noise does not vanish under averaging, since it enters exponentially in Eqs.~(\ref{KLR}) and (\ref{KRL}), making the dynamics intrinsically nonlinear. Representative results for the enantiomeric excess of the three molecular systems considered previously, namely HSOH, H$_2$Se$_2$, and H$_2$Te$_2$, are shown in Fig.~\eqref{fluctuaciones}.
\\
\\

As was previously observed in Refs.~\cite{Hochberg2022PVED,Hochberg2023Noise}, stochastic fluctuations with amplitudes comparable to or smaller than the PVED have only a minor influence on the final enantiomeric excess. In contrast, when the noise amplitude exceeds the PVED scale ($\xi > \epsilon$), significant deviations from the noiseless behavior are obtained. These results suggest that a potential experimental implementation of the proposed mechanism would benefit from averaging over many independent experimental realizations, thereby suppressing stochastic contributions as much as possible. Furthermore, maintaining external noise sources below the PVED scale appears to be a desirable condition for reliably observing the parity-violating effect.
\\
\\
Finally, it is worth discussing other external perturbations that may influence the dynamics independently of the PVED. A particularly relevant example is the presence of external electric or magnetic fields \cite{DeMille2008PV}. Such contributions enter through the off-diagonal terms of the Hamiltonian in Eq.~\eqref{hamilsigmas} and can therefore be interpreted, within the present framework, as corrections to the parameter $\delta$. Unlike the PVED contribution, however, these terms do not appear inside the temperature-dependent exponential factors of Eqs.~(\ref{KLR}) and (\ref{KRL}). Consequently, they are not amplified by the ultralow temperatures of the condensate, so they are expected to play a considerably less significant role within the mechanism considered here.

\section{Conclusions}\label{Sec4}

In this work, we have proposed a mechanism for the indirect detection of the parity-violating energy difference (PVED) in chiral molecules through the observation of a macroscopic enantiomeric excess in an ultracold molecular environment. The model starts from the resonant formation of chiral molecules via ultracold $s$-wave collisions between two achiral diatomic molecules, where the tiny PVED ($\epsilon$) induces a small difference in the resonance energies for the left- and right-handed configurations. This microscopic asymmetry is captured using the Breit-Wigner cross-section and a first-order perturbative expansion, resulting in a maximum asymmetry of $2\epsilon / \Gamma$ at optimal collision energies.

The post-scattering dynamics of the formed chiral molecules are expected to occur in a BEC, leading to the nonlinear interactions characteristic of condensates, alongside typical chiral-molecule effects such as the tunneling effect, the intrinsic PVED, and thermally activated interconversion rates. The nonlinear term leads to an effective bias $\epsilon_{\mathrm{eff}}$ that depends on the population difference between enantiomers, providing a mechanism for amplification. The complete dynamics is described by a set of coupled rate equations that incorporates scattering, tunneling, parity violation, thermal effects, and collective nonlinear interactions.

The central goal of this model is to show how such a system can convert an extremely small parity-violating bias (for now undetectable in conventional high-resolution spectroscopy) into a macroscopic and experimentally observable enantiomeric excess within realistic experimental timescales. Our numerical simulations confirm that this amplification can be achievable on timescales consistent with coherence times reported in optimized ultracold atomic and molecular experiments.

The role played by the temperature in the BEC is crucial: lower temperatures strongly enhance the sensitivity of the interconversion rates to the effective bias, acting as the primary amplification factor. In contrast, the resonance width $\Gamma$ and the initial number of molecules $N_0$ have negligible influence on the final enantiomeric excess, as long as $\Gamma \gg \epsilon$ (to remain in the perturbative regime) and $N_0$ is large enough to support BEC formation. This parameter independence provides considerable experimental robustness and flexibility.

Although BECs of chiral molecules have not yet been realized, all results are based on physically realistic parameters from the literature to ensure that the predicted regime may be experimentally accessible in the future. We find that, for the range of PVED values expected in chiral molecules and their associated tunneling parameters $\delta$, a complete enantiomeric excess can be achieved when the tunneling parameter satisfies $\delta \gtrsim 10$\,Hz, provided that the mean-field nonlinear coupling is sufficiently strong. Even outside this regime, observable enantiomeric excesses can still be obtained. In particular, we have also applied the model to three specific chiral molecules with literature-reported parameters: HSOH ($\epsilon = 1.2 \times 10^{-2}$\,Hz, $\delta = 6 \times 10^{7}$\,Hz), H$_2$Se$_2$ ($\epsilon = 6$\,Hz, $\delta = 3 \times 10^{4}$\,Hz), and H$_2$Te$_2$ ($\epsilon = 90$\,Hz, $\delta = 9 \times 10^{2}$\,Hz). For these systems, complete enantiomeric excess ($ee(t)$ $\approx 1$) is predicted under moderate nonlinear couplings, even when the intrinsic PVED is very small, underscoring the potential power of the proposed amplification mechanism.

We have also considered a possible non-PVED amplification effect, including some results concerning thermal fluctuations and a discussion about other external perturbations such as external electric or magnetic field, which, in principle, are not amplified within the current mechanism. This analysis demonstrates that the nonlinear amplification mechanism does not indiscriminately amplify arbitrarily small stochastic perturbations into a macroscopic signal. Instead, the PVED-induced bias remains robust against fluctuations of comparable magnitude. It is therefore suggested that an experimental implementation should average over independent realizations in order to suppress as much as possible stochastic contributions.

Although molecular BECs of tetraatomic chiral species have not yet been achieved experimentally, the rapid progress in producing, cooling, and trapping ultracold polyatomic molecules makes such systems a realistic prospect in the coming years. The present model, therefore, represents a promising theoretical framework and a potential roadmap for future experiments seeking to detect molecular parity violation.

\section*{Acknowledgements}
D. M. -G. and P. B. acknowledge Generalitat Valenciana through PROMETEO PROJECT CIPROM/2022/13. S. M.-A. acknowledges support of a grant from the Ministry of Science, Innovation and Universities with Ref. PID2023-149406NB-I00

\bibliographystyle{unsrt}
\bibliography{referencias.bib}
\end{document}